\documentclass[twocolumn,showpacs,preprintnumbers,prb,amsmath,amssymb]{revtex4}

\usepackage{bm}
\usepackage{epsfig}

\begin{document}

  \title{Spin-wave theory at constant order parameter}
  
  \author{Marcus Kollar, Ivan Spremo, and Peter Kopietz}
  
  \affiliation{Institut f\"{u}r Theoretische Physik, Universit\"{a}t
    Frankfurt, Robert-Mayer-Strasse 8, 60054 Frankfurt, Germany}
  
  \date{October 25, 2002; updated March 31, 2003}

  \begin{abstract}
    We derive the low-temperature properties of spin-$S$ quantum
    Heisenberg magnets from the Gibbs free energy $G(M)$ for fixed
    order parameter $M$.  Assuming that the low-lying elementary
    excitations of the system are renormalized spin waves, we show
    that a straightforward $1/S$ expansion of $G(M)$ yields
    qualitatively correct results for the low-temperature
    thermodynamics, {\it{even in the absence of long-range magnetic
        order}}.  We explicitly calculate the two-loop correction to
    the susceptibility of the ferromagnetic Heisenberg chain and show
    that it quantitatively modifies the mean-field result.
  \end{abstract}


  \maketitle

  \section{Introduction}\label{sec:intro}
  
  For many years the magnetically ordered state of quantum Heisenberg
  magnets has been studied with the help of the spin-wave
  expansion.\cite{Keffer66} This expansion is usually implemented by
  expressing the components of the spin operator ${\hat{\bm{S}}}_i = (
  \hat{S}^x_i , \hat{S}^y_i , \hat{S}^z_i )$ at lattice site $i$ in
  terms of canonical boson operators $\hat{b}_i$ and
  $\hat{b}_i^{\dagger}$, using either the Holstein-Primakoff
  transformation\cite{Holstein40} or the Dyson-Maleev
  transformation.\cite{Dyson56,Maleev57} For example, the
  Dyson-Maleev transformation for a spin-$S$ ferromagnet is
  \begin{subequations}
    \begin{align}
      \hat{S}^{+}_i
      & = 
      ( 2 S )^{1/2} 
      \left[ 1 - \hat{b}_i^{\dagger} \hat{b}_i/ 2S \right]
      \hat{b}_i\,,
      \label{eq:Splus}
      \\
      \hat{S}^{-}_i
      & = 
      (2 S )^{1/2}
      \hat{b}_i^{\dagger}
      \label{eq:Sminus} 
      \,, 
      \\
      \hat{S}^{z}_i
      & = 
      S - \hat{b}^{\dagger}_i \hat{b}_i
      \label{eq:Sz}
      \,,
    \end{align}
  \end{subequations}
  where $ \hat{S}^{\pm}_i = \hat{S}^x_i \pm i \hat{S}^y_i$, and the
  number of bosons at a lattice site may not exceed $2S$ in order to
  faithfully represent the $2S+1$ eigenstates of $\hat{S}^{z}_i$.  If
  long-range magnetic order is present in the $z$ direction and the
  number of bosons is small compared to $2S$, the two-body bosonic
  interaction arising from the term in Eq.\ (\ref{eq:Splus}) involving
  three bosonic operators can be treated within conventional many-body
  perturbation theory, controlled by the formal small parameter $1/S$.
  Due to the maximum-occupancy constraint the effective bosonic
  Hamiltonian implicitly contains projection operators that eliminate
  the unphysical part of the bosonic Hilbert space.  Usually, these
  projection operators are simply ignored.  For Heisenberg
  ferromagnets in three dimensions it was shown by Dyson\cite{Dyson56}
  that in the thermodynamic limit the low-temperature thermodynamics
  can indeed be obtained without taking into account the so-called
  kinematical interactions associated with these projection operators.
  
  While in the 1960s and 1970s ordered magnets have been intensely
  studied, in recent years the center of attention has shifted to
  low-dimensional magnets without broken symmetries.  In this case the
  conventional spin-wave approach described above is not applicable,
  because it relies on the existence of long-range magnetic order.
  Nevertheless, in many magnetic materials the elementary excitations
  still resemble the spin waves of an ordered magnet.  For example, in
  two-dimensional quantum Heisenberg ferromagnets\cite{Kopietz89} and
  antiferromagnets\cite{Chakravarty89} at low but finite temperatures,
  where the order parameter correlation length $\xi$ is exponentially
  large, spin waves with wave vectors $ | {\bm{k}} | \gg \xi^{-1}$ are
  well-defined elementary excitations.\cite{Kopietz90} Other examples
  for systems where the low-energy physics is dominated by elementary
  excitations of the spin-wave type are Haldane-gap antiferromagnets
  (i.e. one-dimensional Heisenberg antiferromagnets with integer spin
  $S$) and one-dimensional Heisenberg ferromagnets with arbitrary
  spin.
  
  To study the low-temperature properties of these systems, several
  methods have been proposed.  The Schwinger-boson mean-field theory
  of Arovas and Auerbach\cite{Arovas88} is perhaps aesthetically most
  appealing.  However, going beyond the mean-field approximation
  within the Schwinger-boson approach has turned out to be quite
  difficult.\cite{Trumper97} At the mean-field level the modified
  spin-wave theory (MSWT) proposed by Takahashi\cite{Takahashi86} is
  an alternative to the Schwinger-boson approach. MSWT yields results
  that agree with the predictions of Schwinger-boson mean-field
  theory up to numerical prefactors.  This is not surprising, because
  both approaches are in fact equivalent to a one-loop renormalization
  group calculation.\cite{Kopietz89,Chakravarty89} Recently,
  Takahashi's MSWT has also been applied to more complex problems,
  such as frustrated\cite{Pisanova01} or disordered
  magnets,\cite{Wan02} or magnetic molecular
  clusters.\cite{Yamamoto02} However, the MSWT has shortcomings: (i)
  it is very difficult to systematically calculate corrections due to
  interactions between spin waves within MSWT and (ii) the absence of
  long-range magnetic order is not obtained as a result, i.e., the
  magnetization is set to zero by hand; this leads to ambiguity in the
  choice of the constraint if the MSWT is applied to systems with more
  complicated magnetic order, such as ferrimagnets.\cite{Yamamoto98}
  In this work we shall show that these problems can be resolved
  within the conventional spin-wave approach simply by performing the
  calculation at {\it{constant order parameter}}.
  
  This paper is organized as follows. In Sec.~\ref{sec:cop} we discuss
  the calculation of thermodynamic observables at constant order
  parameter.  In Sec.~\ref{sec:swt} this approach is applied to the
  Heisenberg ferromagnet in $D$ $=$ $1,2,3$ dimensions within linear
  spin-wave theory. Hartree-Fock and two-loop corrections are obtained
  for the one-dimensional case in Sec.~\ref{sec:beyond}. The work is
  summarized in Sec.~\ref{sec:concl}.

  \section{Thermodynamics at constant order parameter}\label{sec:cop}
  
  In this section we discuss the calculation of thermodynamic
  observables at constant order parameter.  Although this approach is
  applicable to a variety of correlated systems with order parameter
  $\hat{M}$ and corresponding conjugate field $h$, here we will focus
  on the spin-$S$ Heisenberg ferromagnet with zero-field Hamiltonian
  \begin{align}
    \hat{H} = - J \sum_{ \langle ij \rangle } 
    {\hat{\bm{S}}}_i \cdot {\hat{\bm{S}}}_j 
    \label{eq:Heisenberg}
    \,,
  \end{align}
  where the sum is over all nearest-neighbor pairs of a
  $D$-dimensional hypercubic lattice with $N$ sites, and $ J > 0$ is
  the exchange coupling.  In this case the order parameter is simply
  the total magnetization, given by $\hat{M} = \sum_{i=1}^{N}
  \hat{S}^{z}_i$, and $h$ is the homogeneous magnetic field (in
  suitable units).  Applications to antiferromagnets or
  more complicated magnetic systems are straightforward.
  
  Let us first recall some elementary thermodynamics.  For fixed field
  $h$ and temperature $T$, thermodynamic observables can be obtained
  from the Helmholtz free energy (setting the Boltzmann constant to
  unity)
  \begin{align}
    F (  h ) = - T \ln \mathop{\text{Tr}}
    e^{- ( \hat{H} - h \hat{M}  ) /T }
    \label{eq:freeenergy}
    \,,
  \end{align} 
  where the dependence on $T$ is suppressed for brevity.  Given $F ( h
  )$, the magnetization is obtained as
  \begin{align}
    M (  h ) = 
    - \frac{ \partial F (  h )}{\partial h }
    \,.
    \label{eq:magdef}
  \end{align}
  Alternatively, we may choose to fix the magnetization and adjust the
  magnetic field appropriately.  The corresponding thermodynamic
  potential is the Gibbs free energy $ G ( M )$, which is related to
  the Helmholtz free energy via a Legendre
  transformation,\cite{Negele88}
  \begin{align}
    G (  M ) & =  h(M) M + F (  h (  M ))
    \nonumber
    \\
    & =
    - T \ln \mathop{\text{Tr}}
    e^{-  [ \hat{H} - h(M)  ( \hat{M}  - M ) ]/T }
    \label{eq:Gibbs}
    \,,
  \end{align}
  where the function $h ( M )$ is obtained from Eq.~(\ref{eq:magdef}).
  From $G ( M)$ we obtain the equation of state in the form $h = h ( M
  )$ via
  \begin{align}
    h (  M ) =    \frac{ \partial  G (  M )}{\partial M }
    \,,
    \label{eq:equationofstate} 
  \end{align}
  which shows that the equilibrium magnetization for vanishing field
  is an extremum of $G(M)$.
  If the system has a finite spontaneous magnetization $M_0 = \lim_{ h
    \rightarrow 0^+ } M ( h )$, then the generic expected behavior of
  $G ( M )$ is, for $M \geq M_0$,
  \begin{align}
    G (  M ) = G ( M_0 ) + \frac{(M- M_0 )^2}{2 \chi } 
    +O [  (M - M_0 )^3 ]
    \label{eq:GammaM}
    \,,
  \end{align}
  while for $ | M | <  M_0 $ the Gibbs free energy has the constant
  value $G ( M_0 )$; see, for example, Ref.~\onlinecite{Ma85}.  Here
  \begin{align}
    \chi^{-1}  =  
    \left. \frac{ \partial h ( M )}{\partial M } \right|_{M_0}
    = \left. \frac{ \partial^2 G (  M )}{\partial M^2 } \right|_{M_0}
    \label{eq:susdef}
  \end{align}
  is the inverse longitudinal order parameter susceptibility
  for vanishing external field. These
  expressions are also valid in the absence of spontaneous symmetry
  breaking, where $M_0 = 0$.  Note that, in general, $ G ( M ) = G ( -
  M )$, because the spectrum of $\hat{M}$ is symmetric with respect to
  the origin.

  The parameter $h ( M )$ in Eq.~(\ref{eq:Gibbs}) can be viewed as a
  Lagrange multiplier that enforces the condition of constant
  magnetization. The zero-temperature version of the method outlined
  above has been used previously by Georges and
  Yedidia\cite{Georges91} to study spontaneous symmetry breaking in
  the ground state of the Hubbard model. Note that in the limit $T
  \rightarrow 0$ Eq.~(\ref{eq:Gibbs}) can be written
  as\cite{footnotezero} $ G ( M ) = \langle 0| \hat{G } ( M ) | 0
  \rangle $, where $ | 0 \rangle $ is the ground state of the
  ``free-energy operator'' $\hat{G} ( M ) = \hat{H} - h ( M ) [
  \hat{M} - M ]$.  As shown in Ref.~\onlinecite{Georges91}, the
  expansion at constant order parameter is advantageous for the
  calculation of corrections to the mean-field approximation. In the
  following section we show that for low-dimensional Heisenberg
  magnets without long-range order this method yields reasonable
  results even at the level of linear spin-wave theory; the leading
  fluctuation corrections in $D=1$ are then calculated in
  Sec.~\ref{sec:beyond}.

  \section{Linear spin waves at constant order parameter}\label{sec:swt}
  
  We now calculate $G ( M )$ within linear spin-wave theory, i.e., to
  leading order in $1/S$, assuming that the low-lying elementary
  excitations of the system are renormalized spin waves. In this
  approximation the square brackets in Eq.~(\ref{eq:Splus}) are simply
  replaced by unity, so that the Heisenberg Hamiltonian
  (\ref{eq:Heisenberg}) becomes
  \begin{align}
    \hat{H}_{0} = - D J N S^2 
    + \sum_{ \bm{k}} \epsilon_{\bm{k}}  \hat{b}^{\dagger}_{\bm{k}}
    \hat{b}_{\bm{k}}  
    \,,
    \label{eq:H0}
  \end{align}
  where $\epsilon_{\bm{k}} = 2 D J S ( 1 - \gamma_{\bm{k}} )$, with 
  \begin{align}
    \gamma_{\bm{k}} 
    &=
    D^{-1} \sum_{\mu = 1}^D \cos ( {\bm{k}} \cdot
    {\bm{a}}_\mu )\,.\label{eq:gamma_k}
  \end{align}
  For simplicity, we impose periodic boundary conditions on a
  hypercubic lattice with primitive lattice vectors $ {\bm{a}}_{\mu}$
  and lattice spacing $a = | {\bm{a}}_{\mu} |$.  The momentum sum is
  over the first Brillouin zone and $\hat{b}_{\bm{k}}$ is the lattice
  Fourier transform of $\hat{b}_i$.  The corresponding free energy is
  \begin{align}
    F_{0} (  h ) = 
    - D J N S^2 - h N S  + T \sum_{\bm{k}} \ln
    \left[ 1 - e^{ - ( \epsilon_{\bm{k}} + h )/T }
    \right]
    \,.
    \label{eq:freeenergysw}
  \end{align}
  From Eq.~(\ref{eq:magdef}) we then obtain the usual spin-wave result
  for the magnetization
  \begin{align}
    M (h ) = NS 
    - \sum_{ {\bm{k}} }
    [ e^{ ( \epsilon_{\bm{k}} + h )/T } -1 ]^{-1}
    \,.
  \end{align}

  \subsection{Three-dimensional ferromagnet}
  
\begin{figure}[tb]
    \epsfig{file=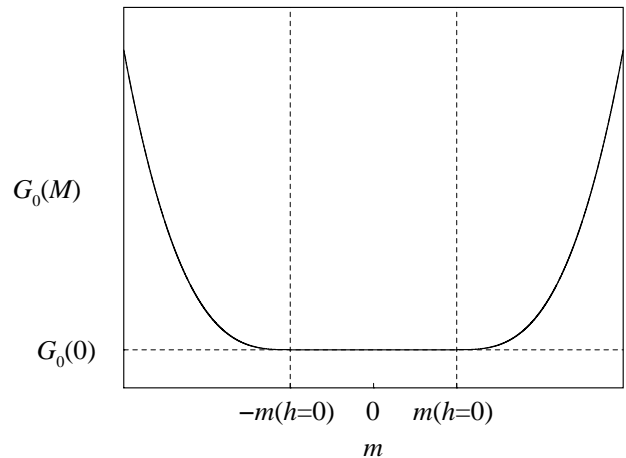,width=6cm,angle=270}
    \caption{Gibbs free energy $G_0 ( M )$ of the three-dimensional 
      Heisenberg ferromagnet within linear spin-wave theory.  Due to
      the divergent longitudinal susceptibility in $D=3$, the Gibbs
      free energy grows cubically for $| m |$ slightly above $m_0
      $, see Eq.~(\ref{eq:Gheis3}).}
    \label{fig:heis3}
  \end{figure}

  It is instructive to begin with linear spin-wave theory for the
  three-dimensional Heisenberg model.  In the thermodynamic limit we
  obtain for the magnetization per site $m = M / N$ to leading order
  in $t = T / JS $ and $v = h/T$
  \begin{align}
    m ( h )  =   S -
    \frac{   \zeta ( \frac{3}{2} )  }{ 8\pi^{3/2} }  t^{3/2}  
    + \frac{ 1  }{4 \pi}  t^{3/2} v^{1/2}
    + O ( t^{5/2} , t^{3/2} v )
    \label{eq:Bloch}
    \,,
  \end{align}
  where $\zeta ( z )$ is the zeta function.  Setting $h =0$ we recover
  the well-known Bloch $T^{3/2}$ law for the spontaneous magnetization
  per site, $m_0 = \lim_{h\to0+}m(h)$, in the ordered state of the
  Heisenberg ferromagnet.  Taking the derivative of
  Eq.~(\ref{eq:Bloch}) with respect to $h$, we see that the
  susceptibility $ \chi = \partial M / \partial h $
  {[Eq.~(\ref{eq:susdef})]} diverges for $h \rightarrow 0$ as
  $h^{-1/2}$.  This divergence of the uniform longitudinal
  susceptibility of a three-dimensional Heisenberg magnet in the
  ordered state is not widely appreciated, although it was noticed
  a long time ago\cite{Holstein40,Dyson56} and has been confirmed by
  renormalization group calculations for the classical Heisenberg
  ferromagnet\cite{Brezin73,Mazenko76} and perturbative calculations
  for the corresponding quantum model.\cite{Vaks68,Kalashnikov80} Due
  to this divergence, the Gibbs free energy $G(M)$ of the Heisenberg
  ferromagnet in $D=3$ does not have the generic
  form~(\ref{eq:GammaM}).  Instead the linear spin-wave result for
  $G(M)$ is, for $m \geq m_0$,
  \begin{multline}
    \frac{G_0 ( M ) }{N  T}  = - \frac{3JS^2}{T}   
    - \frac{ \zeta ( \frac{5}{2} )  }{ 8\pi^{3/2} } t^{3/2}
    + \frac{ 16 \pi^2}{3 t^3 }  ( m - m_0 )^3
    \\
    + O [ (m-m_0 )^4 ]
    \,.
    \label{eq:Gheis3}
  \end{multline}
  From Eq.~(\ref{eq:Bloch}) we note that $h^{1/2}\propto(m - m_0 )$,
  so that we cannot solve for $h$ as a function of $m$ unless $m >
  m_0$.  In light of the above general discussion this is not
  surprising, because for $ |m | <  m_0 $ the Gibbs free energy is
  constant.\cite{Ma85} The behavior of $G_0 ( M )$ as a function of
  $M$ is shown in Fig.~\ref{fig:heis3}.
  The leading $m$ dependence of Eq.~(\ref{eq:Gheis3}) is proportional
  to $( m - m_0 )^3$, which can be traced to the fact that the inverse
  susceptibility vanishes.  By contrast, in $D > 4$ the uniform
  longitudinal susceptibility of the Heisenberg ferromagnet is
  finite,\cite{Brezin73,Mazenko76} so that in this case the Gibbs free
  energy has indeed the generic form (\ref{eq:GammaM}).

  \subsection{One-dimensional ferromagnet}
  
  Let us consider now the one-dimensional case, where we know that the
  Heisenberg ferromagnet does not have any long-range order at any
  finite temperature $T$. In this case the linear spin-wave theory
  result for the magnetization per site is
  \begin{align}
    \frac{ m ( h )}{S}
    =
    1 - \frac{  \zeta ( \frac{1}{2} ) }{2 S \sqrt{\pi} } \sqrt{t}
    - \frac{1}{2S} \sqrt{ \frac{t}{v} }  
    +  O ( t , t^{3/2} v^{-1/2} )
    \label{eq:Bloch1}
    \,,
  \end{align}
  where again $t = T / JS $ and $v = h/T$.  This expression
  predicts a divergent magnetization and susceptibility for
  $h \rightarrow 0$.  However, we can obtain a perfectly
  finite result for the susceptibility at constant
  magnetization [Eq.~(\ref{eq:susdef})]. Solving
  Eq.~(\ref{eq:Bloch1}) for $h$ as a function of $M = N m $
  we obtain
  \begin{align}
    h ( M  ) = \frac{T^2}{ 4 JS [ S - m - 
      \frac{  \zeta ( \frac{1}{2} )  }{2 \sqrt{\pi} }
      \sqrt{t} ]^2 }
    \,.
    \label{eq:hm}
  \end{align}
  According to Eq.~(\ref{eq:susdef})
  this implies for the inverse susceptibility 
  \begin{align}
    {\chi}^{-1}  = \frac{T^2}{2 N JS  [ S - m - 
      \frac{  \zeta ( \frac{1}{2} )  }{2 \sqrt{\pi} }
      \sqrt{t} ]^3 }
    \,.
  \end{align}
  Anticipating that in one dimension $m = 0$, we obtain for the
  susceptibility per site at low temperatures
  \begin{align}
    \frac{ \chi}{N} = \frac{ 2 JS^4}{T^2} \left[ 1 - 
      \frac{3}{S} \frac{  \zeta ( \frac{1}{2} )  }{2 \sqrt{\pi} }
      \sqrt{t} + O ( t ) \right]
    \,.
    \label{eq:chires}
  \end{align}
  This expression agrees exactly\cite{footnotesus} with the
  prediction of the MSWT advanced by Takahashi,\cite{Takahashi86} who
  argued that Eq.~(\ref{eq:chires}) is indeed the correct asymptotic
  low-temperature behavior of the susceptibility for arbitrary $S$.
  For $S = 1/2$ the nearest-neighbor Heisenberg chain is exactly
  solvable via Bethe ansatz,\cite{Takahashi85} so that in this case
  one can obtain an independent check of Eq.~(\ref{eq:chires}).
  Indeed, from a numerical analysis of the Bethe-ansatz integral
  equations\cite{Takahashi85} Takahashi\cite{Takahashi86} found
  perfect agreement with Eq.~(\ref{eq:chires}) for $S = 1 /2$, which
  is remarkable because a priori linear spin-wave theory is only
  expected to be accurate in the ordered state and for large $S$.  We
  shall further comment on this agreement below.

\begin{figure}[tb]
    \epsfig{file=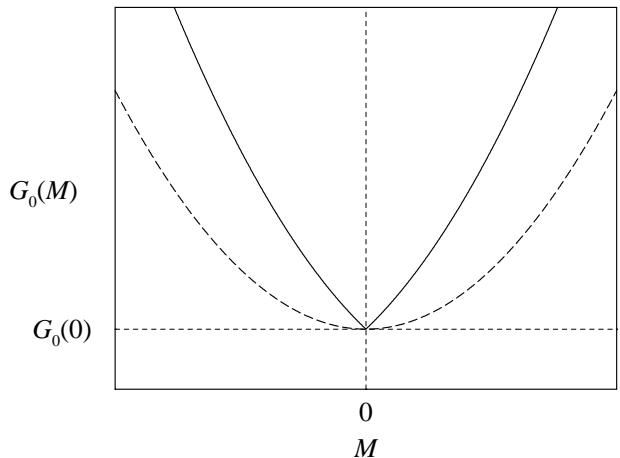,width=6cm,angle=270}
    \caption{
      Solid line: Gibbs free energy $G_0 (M)$ of the one-dimensional
      Heisenberg ferromagnet within linear spin-wave theory, see
      Eq.~(\ref{eq:Gibbsone}). The cusp at $m = 0$ is an unphysical
      artefact of the spin-wave expansion, which is related to the
      neglect of the kinematical interaction between the spin waves.
      The dashed line is the subtracted Gibbs free energy $
      \tilde{G}_0 ( M ) = G_0 ( M ) - h (0) | M |$.}
    \label{fig:heis1}
  \end{figure}

  Within linear spin-wave theory, the Gibbs free energy per site is 
  given by
  \begin{multline}
    \frac{ G_0 ( M ) }{ N T } = -  \frac{J S^2}{T} - 
    \frac{ \zeta ( \frac{3}{2} ) }{ 2 \sqrt{\pi} } \sqrt{t}
    \\
    + \frac{1}{N T} \left[
      h  (0 ) | M |  + \frac{ M^2 }{2  \chi} + O ( | M |^3 )
    \right]
    \label{eq:Gibbsone}
    \,,
  \end{multline}
  where $ \chi$ is given in Eq.~(\ref{eq:chires}) and
  \begin{align}
    h ( 0 ) = \frac{T^2 }{ 4 J S^3} + O ( T^{5/2} )
    \,,
    \label{eq:h0}
  \end{align}
  see Eq.~(\ref{eq:hm}).  In writing Eq.~(\ref{eq:Gibbsone}) we
  have used the fact that our spin-wave calculation yields $G_0 ( M )$
  only for $M \geq 0$ and that the exact $G ( M )$ is an even function
  of $M$.  Note that $G_0 ( M )$ assumes a minimum at $M = 0$,
  indicating the absence of long-range order. However, as shown in
  Fig.~\ref{fig:heis1}, linear spin-wave theory predicts an unphysical
  cusp in the Gibbs free energy at $M = 0$.
  The finite slope $ h ( 0 ) = \partial G_0 / \partial M |_{ M = 0^+}$
  can be identified with the variational parameter $- \mu $ introduced
  by Takahashi,\cite{Takahashi86} which in his calculation plays the
  role of a chemical potential for the Dyson-Maleev bosons, enforcing
  the condition of zero magnetization. On the other hand, it is
  physically clear that for $ T >0 $ any finite value of the external
  field will always be accompanied with a finite magnetization, so
  that an exact calculation of $G ( M )$ should yield $\lim_{M
    \rightarrow 0 } h ( M ) = 0$. Therefore we expect that the exact
  Gibbs free energy in one dimension has the form given in
  Eq.~(\ref{eq:GammaM}).  The cusp of the spin-wave result for the
  Gibbs free energy is related to the fact that in our simple
  spin-wave calculation we have ignored the {\it{kinematical}}
  interaction between the spin waves which arises from the
  maximum-occupancy constraint.\cite{Kollar02} Fortunately, this cusp
  is irrelevant for the calculation of the zero-field thermodynamics,
  which can be also obtained from the subtracted Gibbs free energy
  \begin{align}
    \tilde{G}_0 ( M ) = G_0 ( M ) - h (0 ) | M |
    \,,
    \label{eq:G0sub}
  \end{align}
  see Fig.~\ref{fig:heis1}.  Note that $\tilde{G}_0( M )$ has
  the generic behavior given in Eq.~(\ref{eq:GammaM}), with
  the susceptibility given by Eq.~(\ref{eq:chires}).

  \subsection{Two-dimensional ferromagnet}
  
  For completeness we now discuss the case $D=2$, where the spontaneous
  magnetization of the Heisenberg ferromagnet is zero at any finite
  temperature $T$.  The result of linear spin-wave theory for the
  magnetization is (with $t = T / JS $, $v = h/T$)
  \begin{align}
    m ( h )
    =
    S-\frac{t}{4\pi}
    {[-\ln v
      +\frac{v}{2}
      +\frac{\zeta(2)}{8}t
      +O(t^2,v^2)
      ]}
    \label{eq:Bloch2dim}
    \,,
  \end{align}
  which again diverges for $h\to0$. The function $h(m)$ is
  obtained as
  \begin{align}
    h ( m )
    = 
    Te^{4\pi(S-m)/t}
    [1+O(t)]
    \,,
  \end{align}
  and the result for the susceptibility at $m=0$ is
  \begin{align}
    {\chi}  = 
    \frac{e^{4\pi S/t}}{4\pi JS}
    [1+O(t)]
    \,,
    \label{eq:chitwo}
  \end{align}
  which diverges for $T\to0$. These expressions are
  analogous to Takahashi's results.\cite{Takahashi86}
  Finally, the Gibbs free energy takes the form
  \begin{align}
    \frac{ G_0 ( M ) }{ N T } & =
    -\frac{\zeta(2)}{4\pi}t
    +
    e^{-4\pi S/t}
    \Big(
    \frac{t}{4\pi}
    |m|
    +
    \frac{m^2}{2}
    \Big)
    +
    O(t,m^3)
    \,,
  \end{align}
  i.e., with a minimum at $m=0$, again with an unphysical cusp; in
  this sense the situation is rather similar to that in $D=1$. Note,
  however, that in $D=2$ it is known that a two-loop calculation is
  necessary to obtain the correct low-temperature asymptotics of the
  susceptibility.\cite{Kopietz89} Although the exponential factor
  $\chi \propto \exp [ 4 \pi J S^2 / T ]$ is correctly reproduced by
  mean-field theory, the two-loop correction changes the power of $T$
  in the prefactor of Eq.~(\ref{eq:chitwo}); the correct
  low-temperature behavior of the susceptibility of the quantum
  Heisenberg ferromagnet in two dimensions is $ \chi \propto T^2 \exp
  [ 4 \pi J S^2 / T ]$.  This result is not modified if higher-order
  terms involving more than two loops are
  included.\cite{Kopietz89,Chakravarty89}

  \section{Beyond linear spin-wave theory}\label{sec:beyond}
  
  Because fluctuation effects are usually stronger in lower
  dimensions, we expect that in one dimension the corrections to the
  mean-field result (\ref{eq:chires}) are even more important than in
  $D=2$.  We now explicitly calculate the two-loop correction.  Within
  the Dyson-Maleev formalism the dynamical spin-wave interactions are
  contained in the following two-body Hamiltonian:
  \begin{multline}
    \hat{H}_1  = \frac{D J }{N} 
    \sum_{ {\bm{k}}_1^{\prime} , {\bm{k}}_2^{\prime} , {\bm{k}}_2 , 
      {\bm{k}}_1 }
    \delta_K ( { {\bm{k}}_1^{\prime} + {\bm{k}}_2^{\prime} - 
      {\bm{k}}_2 + {\bm{k}}_1 } )
    \\
    \times   V ( {\bm{k}}_1^{\prime} , 
    {\bm{k}}_2^{\prime} , {\bm{k}}_2 , {\bm{k}}_1 ) 
    \,
    \hat{b}^{\dagger}_{ {\bm{k}}_1^{\prime} }
    \hat{b}^{\dagger}_{ {\bm{k}}_2^{\prime} } \hat{b}_{ {\bm{k}}_2 } \hat{b}_{ {\bm{k}}_1 }
    \,,
    \label{eq:hdmint}
  \end{multline}
  where $\delta_K ( {\bm{k}} )$ denotes momentum conservation modulo a
  reciprocal-lattice vector, and the symmetrized interaction vertex is
  \begin{multline}
    V ( {\bm{k}}_1^{\prime} , 
    {\bm{k}}_2^{\prime} , {\bm{k}}_2 , {\bm{k}}_1 ) =
    - \frac{1}{4} \Bigl[ 
    \gamma_{ {\bm{k}}_1 - {\bm{k}_1^{\prime} } }
    + \gamma_{ {\bm{k}}_1 - {\bm{k}_2^{\prime} } }
    \\
    + \gamma_{ {\bm{k}}_2 - {\bm{k}_1^{\prime} } }
    + \gamma_{ {\bm{k}}_2 - {\bm{k}_2^{\prime} } }
    - 2 \gamma_{ {\bm{k}_1^{\prime} } }
    -2 \gamma_{ {\bm{k}_2^{\prime} } } 
    \Bigr]
    \,,
  \end{multline}
  with $\gamma_{\bm{k}}$ defined in Eq.~(\ref{eq:gamma_k}).  First let
  us estimate the effect of $\hat{H}_1$ within the self-consistent
  Hartree-Fock approximation.  We write our spin-wave Hamiltonian as $
  ( \hat{H}_0 + \delta \hat{H}_1 ) + ( \hat{H}_1 - \delta \hat{H}_1 )$
  and choose the one-body Hamiltonian $\delta \hat{H}_1$ such that the
  thermal expectation value of the residual interaction $ \hat{H}_1 -
  \delta \hat{H}_1 $ in the ensemble defined by the Hartree-Fock
  Hamiltonian $ \hat{H}_0 + \delta \hat{H}_1$ vanishes.  We obtain
  \begin{align}
    \delta \hat{H}_1 = \sum_{ {\bm{k}} } \Sigma_{ 1} ( {\bm{k}} )
    \hat{b}^{\dagger}_{ {\bm{k}} } \hat{b}_{ {\bm{k}} }  
    - \frac{ 2DJ}{N} \sum_{ {\bm{k}} , {\bm{k}}^{\prime} } V ( {\bm{k}} , 
    {\bm{k}}^{\prime} , {\bm{k}}^{\prime} , {\bm{k}} ) \,
    n_{ {\bm{k}} } \, n_{ {\bm{k}}^{\prime} }
    \,,
    \label{eq:HHF}
  \end{align}
  where ${n}_{ {\bm{k}} } = [ e^{ ( E_{\bm{k}} + h )/T } -1 ]^{-1}$ is
  the thermal occupation of the Hartree-Fock magnon states with
  momentum ${\bm{k}}$, and the Hartree-Fock self-energy is given by
  \begin{align}
    \Sigma_{ 1} ( {\bm{k}} ) = \frac{ 4DJ}{N} \sum_{ {\bm{k}}^{\prime} }
    V ( {\bm{k}} , 
    {\bm{k}}^{\prime} , {\bm{k}}^{\prime} , {\bm{k}} ) \,
    n_{ {\bm{k}}^{\prime} }
    \,.
    \label{eq:sigmaHF} 
  \end{align}
  After some standard manipulations we obtain for the Helmholtz free
  energy within self-consistent Hartree-Fock approximation
  \begin{multline}
    F_{1} (  h ) =
    - D J N S^2 - h N S  + T \sum_{\bm{k}} \ln
    \left[ 1 - e^{ - ( E_{\bm{k}} + h )/T }
    \right]
    \\
    +  DJ N S^2 ( 1 - Z )^2
    \,.
    \label{eq:freeenergy1}
  \end{multline}
  Here $E_{\bm{k}} = Z  \epsilon_{ {\bm{k}} }$,
  and the dimensionless renormalization factor $Z$ satisfies 
  the self-consistency condition
  \begin{align}
    Z = 1 - \frac{1}{NS} \sum_{ {\bm{k}} } (1 - \gamma_{ {\bm{k}} } ) 
    {n}_{ {\bm{k}} }
    \,.
  \end{align}
  The quantity $ZS$ corresponds to the second variational parameter
  $S^{\prime}$ introduced by Takahashi.\cite{Takahashi86} Note that he
  gives a different sign for the last term in
  Eq.~(\ref{eq:freeenergy1}).  In one dimension $ 1 -Z = O ( T^2 )$ at
  low temperatures,\cite{Takahashi86} so that for the calculation of
  the first two terms in low-temperature expansion of thermodynamic
  observables it is sufficient to set $ Z =1 $. We conclude that at
  the Hartree-Fock level the dynamical interaction between spin waves
  does not contribute to the low-temperature asymptotics in $D=1$. At
  this level of approximation our theory is equivalent to MSWT.
  
\begin{figure}[tb]
    \epsfig{file=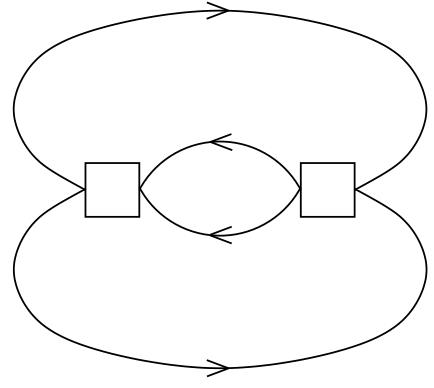,width=5cm,angle=270}
    \caption{
      Feynman diagram describing the leading fluctuation correction to
      the free energy of the ferromagnetic Heisenberg model, see
      Eq.~(\ref{eq:Wmatrix}).  The solid arrows denote the
      Hartree-Fock magnon propagators and the squares are the
      Dyson-Maleev vertices.}
    \label{fig:feynman}
  \end{figure}

  Within our approach it is now straightforward to study spin-wave
  interactions beyond the Hartree-Fock approximation. Therefore we
  simply expand the Helmholtz free energy $F ( h )$ to higher order in
  the interaction and then perform a Legendre transformation to obtain
  the corresponding Gibbs free energy.  We now calculate the first
  fluctuation correction to $F ( h )$. The relevant Feynman diagram is
  shown in Fig.~\ref{fig:feynman}.
  In this approximation the Helmholtz free energy is $ F_2 ( h ) = F_1
  ( h ) + \delta F_2 ( h )$, where $F_1 ( h )$ is given in
  Eq.~(\ref{eq:freeenergy1}) and
  \begin{multline}
    \delta F_2 ( h )  = 
    2 \left( \frac{D J}{N} \right)^2
    \sum_{ {\bm{k}}, {\bm{k}}^{\prime} , {\bm{q}} }
    \frac{ W ( {\bm{k}} , {\bm{k}}^{\prime}  , {\bm{q}}  ) }{
      E_{ {\bm{k}} } + E_{ {\bm{k}}^{\prime}} - E_{ {\bm{k}} + {\bm{q}} }
      - E_{ {\bm{k}}^{\prime} - {\bm{q}} } }
    \\
    \times   \bigl[   
    ( 1 + {n}_{ \bm{k}} ) 
    ( 1 + {n}_{ {\bm{k}}^{\prime} } ) {n}_{ {\bm{k}}
      + {\bm{q}} } {n}_{ {\bm{k}}^{\prime} - {\bm{q}} } 
    \\
    - 
    {n}_{ \bm{k}}   {n}_{ {\bm{k}}^{\prime} }  
    (1 + {n}_{ {\bm{k}}
      + {\bm{q}} })  ( 1 + {n}_{ {\bm{k}}^{\prime} - {\bm{q}} } )
    \bigr]
    \label{eq:F2}
    \,,
  \end{multline} 
  where 
  \begin{multline}
    W ( {\bm{k}} , {\bm{k}}^{\prime}  , {\bm{q}}  )
    =  V ( {\bm{k}} , {\bm{k}}^{\prime} , {\bm{k}} + {\bm{q}} , 
    {\bm{k}}^{\prime} - {\bm{q}} )
    \\\times
    V ( {\bm{k}} + {\bm{q}} , 
    {\bm{k}}^{\prime} - {\bm{q}} ,  {\bm{k}} , {\bm{k}}^{\prime}   )
    \,.
    \label{eq:Wmatrix}
  \end{multline}
  At low temperatures, we may replace the Dyson-Maleev vertex by its
  long-wavelength limit, which in $D$ dimensions is given by $ V (
  {\bm{k}}_1^{\prime} , {\bm{k}}_2^{\prime} , {\bm{k}}_2 , {\bm{k}}_1
  ) \sim - ({\bm{k}}_1 \cdot {\bm{k}}_2) a^2 / 2D$.  For the rest of
  this work we shall explicitly set $D=1$. Then the leading behavior
  of $\delta F_2 ( h )$ for small $t = T / JS$ and small $v = h /T$
  can be calculated analytically.  We obtain for $N \rightarrow
  \infty$
  \begin{align}
    \frac{\delta F_2 ( h ) }{TN}  
    = \frac{1}{16} \frac{ t^{3/2}}{ (2S)^2 v^{1/2}}
    + O ( t^{3/2} , t^{1/2} v )
    \,.
    \label{eq:F2res}
  \end{align} 
  The resulting equation of state is
  \begin{multline}
    \frac{m ( h )}{S}  
    =   1 - \frac{  \zeta ( \frac{1}{2} ) }{2 S \sqrt{\pi}  } \sqrt{t}
    - \frac{1}{2S} \sqrt{ \frac{t}{v} }
    \\
    + \frac{1}{16 }  \left[   \frac{1}{2S} \sqrt{ \frac{t}{v} } \right]^3
    +  O ( t , t^{3/2} v^{-1/2} )
    \label{eq:Bloch2}
    \,.
  \end{multline}
  Comparing this result with the corresponding expression obtained
  within linear spin-wave theory given in Eq.~(\ref{eq:Bloch1}), we
  see that the two-loop correction gives rise to an additional term
  proportional to the third power of $(2 S )^{-1} (t / v)^{1/2} $. But
  linear spin-wave theory predicts that this parameter is actually
  close to unity, as is easily seen by setting $m=0$ in
  Eq.~(\ref{eq:Bloch1}). Hence, {\it{the leading fluctuation
      correction to the Hartree-Fock theory is not controlled by a
      small parameter}}. Note that the extra power of $S^{-1}$ that
  appears in the two-body part of the effective boson Hamiltonian is
  canceled by the singular $h$ dependence of the two-loop correction.
  If we nevertheless truncate the expansion at the two-loop order, we
  obtain from Eq.~(\ref{eq:Bloch2}) for the leading low-temperature
  behavior of the susceptibility,
  \begin{align}
    \frac{\chi}{N} \sim C_{\chi}  \frac{J S^4 }{ T^2}
    \,,
    \label{eq:chitrue}
  \end{align}
  with $C_{\chi} \approx 1.96$, which is slightly smaller than the
  linear spin-wave prediction $C_{\chi} = 2$, and significantly
  smaller than the result $C_{\chi} = 3$ obtained within
  Schwinger-boson mean-field theory.\cite{Arovas88} We suspect that
  corrections involving more loops will involve higher powers of the
  parameter $(2 S )^{-1} ( t/v)^{1/2}$ in Eq.~(\ref{eq:chitrue}),
  which give rise to additional finite renormalizations of $C_{\chi}$.
  Hence, a numerically accurate expression for the low-temperature
  susceptibility of a one-dimensional Heisenberg ferromagnet cannot be
  obtained from a truncation of the $1/S$ spin-wave expansion at some
  finite order.  Note that quantum Monte Carlo simulations for the $S
  =1/2$ nearest-neighbor Heisenberg chain\cite{Kopietz89b} give
  $C_{\chi} = 1.58 \pm 0.03$, supporting the scenario described above.
  In light of these results it is puzzling that from the numerical
  analysis of the Bethe-ansatz integral equations for $S = 1/2$
  Takahashi\cite{Takahashi85,Takahashi86} obtained $C_{\chi} = 2$.
  Possibly this is related to difficulties in extracting the true
  asymptotic low-temperature behavior of the susceptibility from the
  Bethe-ansatz integral equations.\cite{Takahashi86,Schlottmann85}
 
  Although at first sight our spin-wave expansion for the
  susceptibility appears to be controlled by the small parameter
  $1/S$, this parameter is renormalized by the infrared singularity of
  the two-loop correction to the mean-field result.  This phenomenon
  is familiar from the weak-coupling calculation of the two-loop
  correction to the ground-state energy of the repulsive Hubbard model
  at constant staggered magnetization in one and two
  dimensions.\cite{Georges91,Kopietz93} Due to infrared singularities
  inherent in the loop integrals, those expansions are effectively in
  powers of the Hubbard interaction $U$ multiplied by a function of
  the order parameter; as a consequence the two-loop correction to the
  ground-state energy has the same order of magnitude as the
  Hartree-Fock term.

  \section{Conclusions}\label{sec:concl}
  
  In summary, we have shown that conventional spin-wave expansion at
  constant order parameter is an alternative to Takahashi's modified
  spin-wave theory, which nowadays is one of the most popular
  mean-field methods to study low-dimensional magnets without
  long-range order.  Due to the conceptual simplicity of our method,
  we can systematically calculate corrections to the mean-field
  approximation using conventional diagrammatic methods.  We have
  explicitly calculated the leading fluctuation correction to the
  mean-field result for the susceptibility of the ferromagnetic
  Heisenberg chain.  We have found that in one dimension the
  predictions of MSWT are at most qualitatively correct, because
  fluctuation corrections are not controlled by a small parameter.
  
  Furthermore, compared to MSWT the present approach has the
  conceptual advantage that the introduction of Lagrange multipliers
  by hand is not necessary, since their role is played by external
  fields instead.  The difference between these approaches is best
  visible for systems with more complicated order parameters.  For
  example, recent work on molecular magnets\cite{Yamamoto02} and
  ferrimagnets\cite{Yamamoto98} has shown that the formulation of MSWT
  for such sytems is difficult and that the proper choice of
  constraints is not clear a priori. On the other hand, spin-wave
  theory at constant order parameter naturally yields the absence of
  long-range order in low dimensions \emph{as a result}; applications
  to antiferromagnets and ferrimagnets, with vanishing homogeneous and
  staggered magnetizations, are in progress.\cite{Kollar02}
  
  This work was supported by the DFG via Forschergruppe FOR 412,
  Project No. KO 1442/5-1.


\begin{thebibliography}{99}
\bibitem{footnotezero}
    Of course, for a ferromagnet the order parameter and the Hamiltonian
    commute, so that the zero-temperature limit is trivial due to the
    absence of quantum fluctuations.
\bibitem{footnotesus}
    Note that the normalization of the susceptibilty in
    Ref.~\onlinecite{Takahashi86} is different from ours, so that we
    should identify $\chi^{\rm Takahashi} = \frac{4}{3 } \chi /N$.
    Taking this into account, our Eq.~(\ref{eq:chires}) agrees exactly
    with the susceptibility obtained by Takahashi.
\end{thebibliography}
\end{document}